# Observation of Conductance Quantization in InSb Nanowire Networks

Elham M. T. Fadaly [1,2 ‡], Hao Zhang [1 ‡], Sonia Conesa-Boj [1], Diana Car [1,2], Önder Gül [1],

Sébastien R. Plissard [2§], Roy L. M. Op het Veld [1,2], Sebastian Kölling [2], Leo P. Kouwenhoven [1,3],

Erik P. A. M. Bakkers [1,2 *]

[1] *QuTech and Kavli Institute of Nanoscience, Delft University of Technology, 2600 GA Delft, The Netherlands*

[2] *Department of Applied Physics, Eindhoven University of Technology, 5600 MB Eindhoven, The Netherlands*

[3] *Microsoft Station Q Delft, 2600 GA Delft, The Netherlands*

[§] *Present address: NRS-Laboratoire d'Analyze et d'Architecture des Systemes (LAAS),*

*Universitéde Toulouse, 7 avenue du Colonel Roche, F-31400 Toulouse, France.*

[‡]These authors contributed equally to this work

Email: elham.fadaly@gmail.com; e.p.a.m.bakkers@tue.nl

## ABSTRACT

Majorana Zero Modes (MZMs) are prime candidates for robust topological quantum bits, holding a great promise for quantum computing. Semiconducting nanowires with strong spin orbit coupling offers a promising platform to harness one-dimensional electron transport for Majorana physics. Demonstrating the topological nature of MZMs relies on braiding, accomplished by moving MZMs around each other in a certain sequence. Most of the proposed Majorana braiding circuits require nanowire networks with minimal disorder. Here, the electronic transport across a junction between two merged InSb nanowires is studied to investigate how disordered these nanowire networks are. Conductance quantization plateaus are observed in all contact pairs of the epitaxial InSb nanowire networks; the hallmark of ballistic transport behavior.

Semiconductor nanowires (NWs) with strong spin-orbit coupling, e.g. InSb or InAs, provide a promising platform to study Majorana-related physics. Majorana Zero Modes (MZMs) are predicted to appear in carefully engineered solid state systems, which can be utilized as building blocks for topological quantum computer by using their non-abelian properties.[1–3] When a 1D semiconductor with low disorder and strong spin-orbit interaction (SOI) is brought into contact with a superconductor under an external magnetic field, MZMs appear at both ends of the proximitized semiconductor region.[4,5] Signatures of MZMs have been experimentally observed in InSb and InAs semiconductor/superconductor hybrid NW systems.[6–14] The ambitious goal of performing logical operations utilizing Majoranas and studying their non-Abelian properties, i.e. braiding, requires NW networks. These networks are required to move MZMs around each other or to tune the coupling between each other in a certain sequence, according to most of the theoretical proposals.[15–19] The semiconductor NW network system of our preference is InSb since it exhibits a higher electron mobility,[20] stronger SOI,[21] and larger Landé g-factor than InAs.[22–25]

High structural quality InSb NW networks have been recently reported by Car et al.[26] To date, several transport experiments have been performed on branched NW networks.[27–30] However, one remaining unanswered question is the degree of disorder in this system. Disorder is a serious issue as it can mimic Majorana signatures or even can be completely detrimental to the topological protection of MZMs.[9] In InSb NW devices, disorder has been diminished to the point where ballistic transport, whose hallmark is quantized conductance plateaus, is observed.[25,31] Yet, ballistic transport in NW networks has not been reported yet.

Here, we study the electronic transport across the junction between two merged InSb NWs using the optimized nanofabrication recipe reported in ref [9] and [25]. In this work, the evolution of the quantized conductance plateaus in each contact pair is studied as a function of magnetic field, gate voltage and source-drain bias voltage. Quantized conductance plateaus are observed consistently in all contact pairs of the epitaxially grown InSb NW networks in three different devices. The Landé- g factor of the first subband is extracted for each contact pair in the NW network. Additionally, the structural quality of the junction between the crossed NWs in a representative device has been inspected by cross-sectional transmission electron microscopy (XTEM) and correlated with the transport measurements.

The InSb NW networks used in this work have been synthesized by gold (Au)-catalyzed Vapor-Liquid-Solid (VLS) growth mechanism in a Metal Organic Vapor Phase Epitaxy (MOVPE) machine. Kinked InP NWs grown on InP <001> substrate have been used as stems for the growth of InSb <111> B NW networks. Since the InSb NW networks are in an epitaxial relationship to the InP substrate, the crossed InSb NWs meet under an angle of 109.5° which corresponds to the crystallographic angle between two <111> B directions in a zinc blende crystal structure. Further details related to the NW networks growth and the structural quality of these networks have been reported in ref [26]. After growth, the NW networks are deterministically transferred from the growth chip via a nanomanipulator in a scanning electron microscope (SEM) to the desired chip for device fabrication as shown in the **supporting information (SI)-1 and SI-2**. **Figure. 1a** shows a false color 30-degree tilted SEM image of a typical InSb NW network device after the fabrication process. The InSb network is deposited on a P++-doped Si substrate covered with 285 nm of $SiO_2$. The electrical contacts to the InSb network are defined by electron beam lithography followed by the evaporation of metallic normal contacts (10/210 nm of Cr/Au). Prior to contact deposition, the surface oxide of the InSb network is removed via sulfur passivation and short in-situ He- ion milling. Further details on the device fabrication process are explained in **SI** in the fabrication recipe section. Afterwards, the sample is mounted in a $^{3}$He cryostat with a base temperature of 300 mK, single axis magnet of 9 T and measured using a standard lock-in technique at 73 Hz with an excitation route mean square voltage ($V_{RMS}$) 30 μV. A back gate voltage ($V_{gate}$) is applied to the Si substrate and the $SiO_2$ acts as a gate dielectric. A bias voltage is applied to one contact and current is measured through another grounded contact, while the rest of the contacts are left floating. This setup is the same for each contact pair in the NW network as illustrated schematically in **Figure. 1b**. All data reported below are calibrated after subtracting resistances of the filters and the wiring of the cryostat system.

Differential conductance is measured using a standard lock-in technique by applying a small ac excitation voltage ($V_{ac}$) at a fixed dc bias voltage and measuring the ac current ($I_{ac}$) such that $G = dI_{ac}/dV_{ac}$. The channel length denoted by the contact spacing between the different contact pairs is measured from the top-view SEM image of the device shown in **Figure. 1a**. The lengths of the two straight contact pairs labelled as A-C and B-D are 770 and 640 nm, respectively. The lengths of the kinked contact pairs A-B, A-D, B-C, and D-C are 700, 610, 700, and 700 nm, respectively. In principal, ballistic transport behavior is very challenging to realize in such a NW network with this geometry for the following reasons. NWs system already has a large surface to volume ratio which can cause electrons to scatter back to the source reservoir. In addition, our NW network devices has long contact spacing between the contact pairs which is much larger the electrons mean free path reported in single ballistic NW devices. [25,31,32]. Concerning the geometry of the NW network devices, the interface between merged NWs can induce additional scattering. Furthermore, electrons traversing the crossed networks are required to follow a bended trajectory to comply with the device geometry.

**Figure. 2(a-d)** shows differential conductance ($G = dI/dV_{\text{bias}} = I_{\text{AC}}/V_{\text{AC}}$) of three different contact pairs of the network (A-B, A-C, A-D) as a function of $V_{gate}$ and $B$ at $V_{bias} = 0\ mV$ (the data for B-C, B-D, C-D can be found in **SI-5** and **SI-6**). For increased clarity, line cuts at different $B$ value; 0 T(green), 5T (red), and 8T (black) are shown in the bottom panels with a horizontal offset. At zero magnetic field, no quantized conductance plateaus are observed which is expected due to the many possible scattering sources previously discussed. As the magnetic field is increased, back scattering of electrons is suppressed and the first spin resolved conductance plateau ($G = e^2/h$ ) is revealed. It becomes more pronounced and flatter at higher magnetic fields. In some devices (**SI- 7**), higher plateaus (up to three plateaus ($e^2/h$, $2e^2/h$, $3e^2/h$)) are observed. In this paper, the focus is on the first plateau which is present in all our devices. The line cuts in the bottom panels of the color plots in **Figure. 2** exhibit a clear evolution of the first quantized plateau towards ($G = e^2/h$) with increasing $B$ confirming that magnetic field helps on ballistic transport by suppressing electrons back scattering. All contact pairs in the NW network show extended conductance plateaus around ($e^2/h$) except for one straight channel (A-C) that exhibits an unexpectedly reduced plateau value (around 0.3 $e^2/h$). Since this straight channel (A-C) is the top wire (in the network that has one end lifted from the dielectric surface), it might be a geometrical effect, i.e the distance between the top wire and the dielectric that varies along the channel, see **SI-3**. The fact that all the other contact pairs involving contact A or C (e.g. A-B, A-D, C-B, C-D) show regular plateau values (around $e^2/h$) proves that the low plateau value of the A-C contact pair is not due to poor electrical contacts of A or C.

For further investigations, at $B = 8.5\ T$, the differential conductance for different contact pairs in the NW network is measured as a function of $V_{gate}$ and $V_{bias}$ as illustrated in the color plots in **Figure. 3(a-c)**. The color plots exhibit diamond shaped regions of constant conductance (G = $e^2/h$) highlighted by dotted black lines. Line cuts of these color plots, indicated by the green dotted line, along ($V_{bias} = 0\ mV$) are shown in the bottom panels. In the middle of the diamond, at zero bias voltage, a prolonged conductance plateau appears at $e^2/h$. This can be explained by the energy spectrum shown in **Figure. 3d** where the chemical potentials of the source and drain ($\mu_s, \mu_d$) are aligned together between $E_{1\uparrow}$ and $E_{1\downarrow}$. At the tips of the diamond where the dotted lines cross each other, the bias voltage is equivalent to the first spin-split subband spacing ($E_{1\downarrow} - E_{1\uparrow} = eV_{bias}$) enabling the extraction of the subband spacing as illustrated schematically in the energy spectrum in **Figure. 3e**. Accordingly, the Landé g-factor ($g_1$) of the first channel for each contact pair in the NW network can be extracted since the measured subband spacing is purely due to Zeeman splitting ($E_{Zeeman} = g\mu_B B$). Using this approach, we estimate $g_1$ of 43, 43, 52 corresponding to the sub band spacing, $\Delta_{subband}$, of 21, 21 and 25 meV for the different channels A-B, A-C and A-D, respectively. The estimated $g_1$ values for the different contact pairs is close to the expected InSb bulk value of 51. Data from additional devices are included in **SI-(7-10)** demonstrating consistent results and ballistic transport in all three devices.

To gain more insight about the structural quality of the junction between crossed NWs in the measured NW networks, device II (its transport data is shown in **SI-7 and SI-8**) was sliced open using focused ion beam (FIB). A 50-nm thin lamella along one of the NWs (indicated by a red line in **Figure. 4a**, was prepared and inspected in a transmission electron microscope (TEM). Both the cross sectional TEM images (**Figure. 4c**) and the energy-dispersive X-ray spectroscopy (EDX) map (**Figure. 4e**), show a sharp and oxide-free interface between the crossed NWs. Moreover, the zinc blende crystal structure of the junction is confirmed by the FFT pattern of the lattice imaged along the [121] direction as shown in **Figure. 4d**. The high quality of the junction based on this structural analysis supports the ballistic transport behavior observed in these devices.

In summary, we have demonstrated conductance quantization between all the contact pairs in InSb NW networks at non-zero magnetic field and this observation is consistent over three different devices. Bias voltage spectroscopy on these quantized plateaus at finite magnetic field measures spin-resolved subband spacing, enabling an estimation of Landé g-factor above 40 in these NW networks. Our results show that our InSb NW networks is a low disorder system which is promising for implementation in future disorder-free Majorana braiding circuits.

**FIGURES**

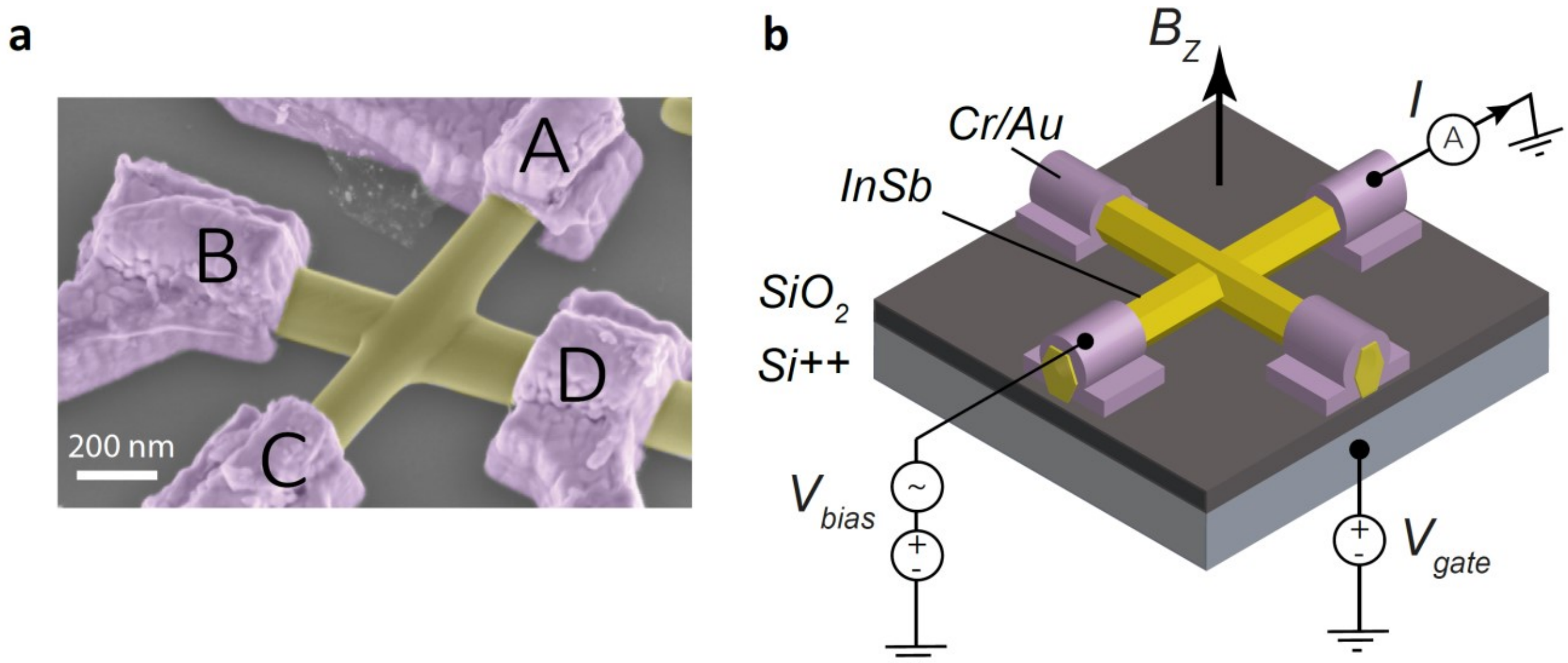

**Figure 1. A typical InSb NW network device:** (a) a false-colored, 30°-tilted SEM image of an InSb NW cross (yellow) deposited on a p++ - doped Si substrate (light grey) covered with 285 nm of $SiO_2$ (dark grey) and contacted with 10/210 nm of Cr/Au (purple). The NW network terminals are labelled A, B, C, and D. (b) schematic illustration of the experimental setup. The Si substrate acts as a global back gate and the $SiO_2$ is the gate dielectric. A gate voltage ($V_{gate}$) is applied to the Si substrate. A source-drain voltage bias ($V_{bias}$) is applied between two terminals of the NW network and the current is monitored between these two terminals. The rest of the terminals are floated. All the measurements are performed at a temperature of 300 mK and out-of-plane magnetic field.

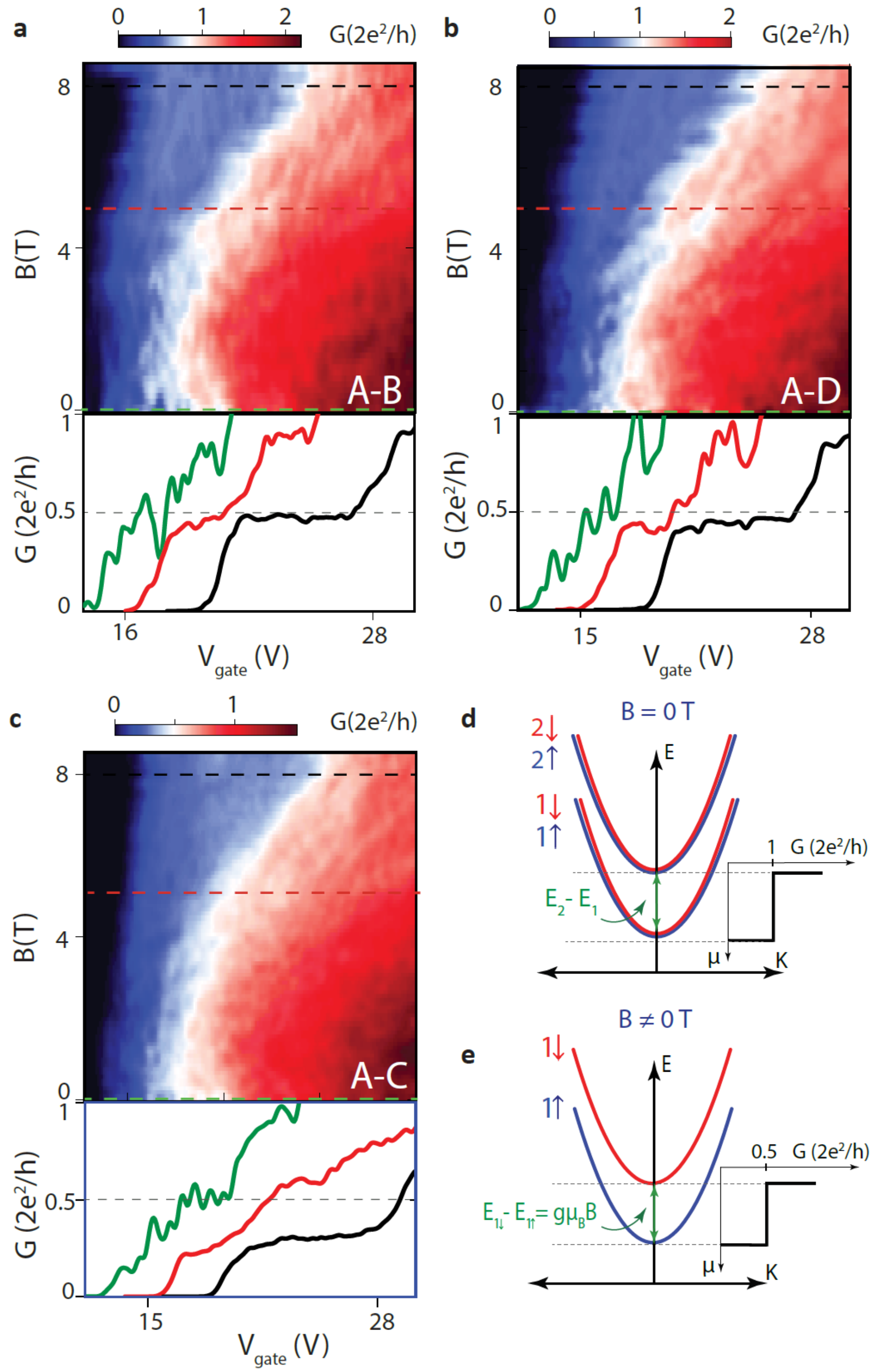

**Figure 2. Evolution of conductance in magnetic field:** (a-c) Color plots of conductance (G = d$I$/d$V_{\text{bias}}$) as a function of $V_{\text{gate}}$ and magnetic field $B$ at $V_{\text{bias}}$ = 0 mV for different contact pair combinations: (a) A-B, (b) A-D, (c) A-C. The bottom panels in (a-c) show I-V traces indicating

line cuts of (a-c) at different $B$ values 0T (green), 5T (red), and 8T (black) with a horizontal offset between the individual traces for clarity. (d-e) energy spectra at different $B$ values, sketching the evolution of the spin resolved subbands. In this panel, the energy spectrum near the bottom of the first subband and the corresponding conductance region are shown. (e) in the absence of magnetic field ($B = 0$), the energy spectrum shows the spin-degenerate first two subbands and the energy spacing between them is denoted as $E_1 - E_2$ where $G = 2e^2/h$. (f) at non-zero magnetic field ($B \neq 0$), the spectrum consists of the spin-split of the first subband ($E_{1\downarrow}, E_{1\uparrow}$) and the energy spacing is due to Zeeman splitting ($\Delta E_{subband} = E_{1\uparrow} - E_{1\downarrow} = g\mu_B B$). When the chemical potential reaches the lowest spin-split subband, the conductance is $e^2/h$ instead of $2e^2/h$.

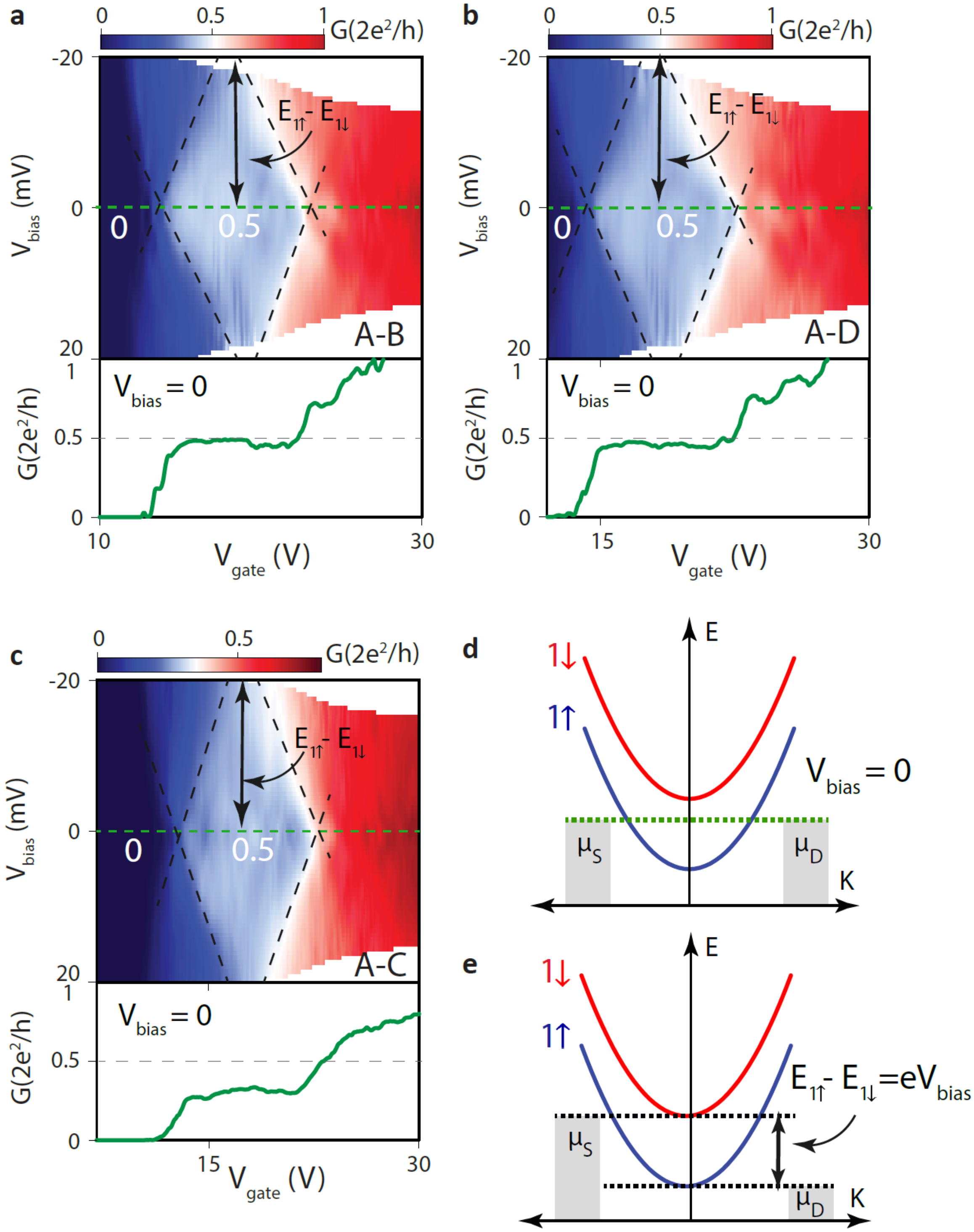

**Figure 3. Voltage bias spectroscopy**: (a-c) color plots of the differential conductance $G = dI/dV_{\mathrm{bias}}$ as a function of $V_{\mathrm{bias}}$ and $V_{\mathrm{gate}}$ at $B = 8.5$ T. A line cut along $V_{\mathrm{bias}} = 0$ mV (green) is shown in the bottom panel. Black dotted lines surrounding a diamond shaped region, indicating the edge of the first quantized conductance plateau, are drawn as guide to the eye. (d-e) energy spectra showing

the spin-split first subband such that (d) at $V_{\text{bias}}$ = 0 mV along the dotted green line, the source and drain chemical potential ($\mu_s$, $\mu_d$) are aligned together in between $\boldsymbol{E_{1\uparrow}}$ and $\boldsymbol{E_{1\downarrow}}$.and (e) at the tip of the diamond, the subband spacing is equivalent to the bias voltage ($\Delta\boldsymbol{E_{subband}} = \boldsymbol{E_{1\uparrow}} - \boldsymbol{E_{1\downarrow}} = \boldsymbol{eV_{bias}}$).

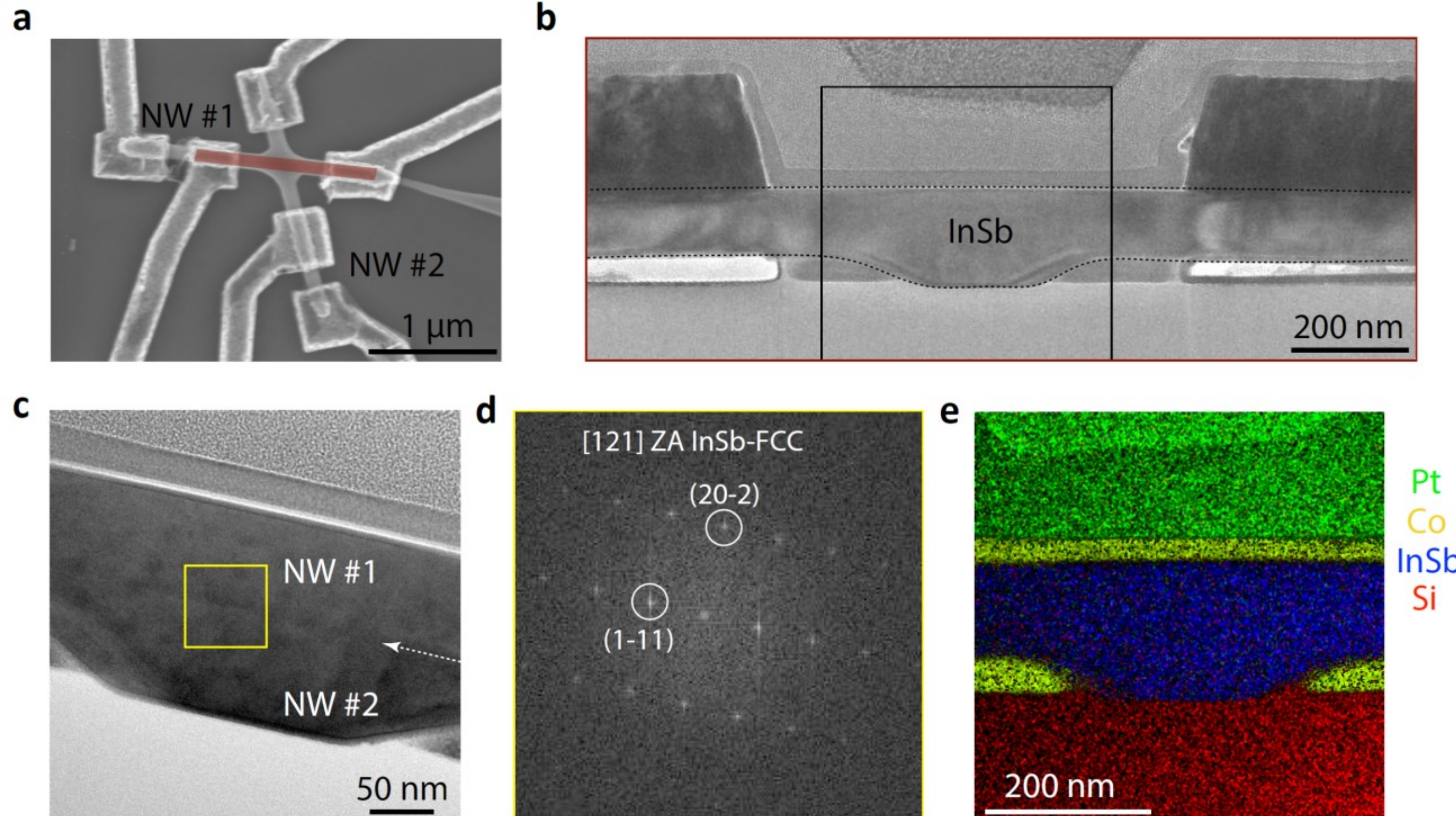

**Figure 4. Structural characterization:** (a) A top-view SEM image of the InSb NW network device. The red rectangle indicates a lamella of the device along one of the nanowires which was prepared using a focused ion beam to be inspected in TEM. The resulting low-magnification bright-field TEM image is shown in (b). The intersection of the two constituent nanowires denoted as NW#1 and NW#2, is highlighted by a black square. (c) High-resolution TEM image of the junction viewed along the [121] zone axis. The white arrow indicates the sharp clean interface between the two crossed nanowires. (d) Fast Fourier transform (FFT) pattern evaluated at the interface between the two crossed nanowires indicated by a yellow square in (c). (e) EDX compositional map of the device cross section. The InSb NW network is shown in blue and Si/$SiO_2$substrate in red. Layers of Pt and Co (shown in green and yellow, respectively) have been deposited during focused ion beam sample preparation to protect the junction from induced damage.

**ACKNOWLEDGMENTS**

The authors thank Kun Zuo, Alain Dijkstra and Ghada Badawy for the critical reading of the manuscript. This work has been supported by The Netherlands Organization for Scientific Research (NWO), Foundation for Fundamental Research on Matter (FOM), European Union Seventh Framework Programme, European Research Council (ERC), Office of Naval Research (ONR) and Microsoft Corporation Station Q.

# Supporting Information: Observation of Conductance Quantization in InSb Nanowire Networks

Elham M. T. Fadaly [1,2], Hao Zhang [1], Sonia Conesa-Boj [1], Diana Car [1,2], Önder Gül [1], Sébastien R. Plissard [2§], Roy L. M. Op het Veld [1,2], Sebastian Kölling [2], Leo P. Kouwenhoven [1,3], Erik P. A. M. Bakkers [1,2 *]

[1] *QuTech and Kavli Institute of Nanoscience, Delft University of Technology, 2600 GA Delft, The Netherlands*

[2] *Department of Applied Physics, Eindhoven University of Technology, 5600 MB Eindhoven, The Netherlands*

[3] *Microsoft Station Q Delft, 2600 GA Delft, The Netherlands*

[§] *Present address: NRS-Laboratoire d'Analyze et d'Architecture des Systemes (LAAS), Universitéde Toulouse, 7 avenue du Colonel Roche, F-31400 Toulouse, France.*

[*] Email: Elham.fadaly@gmail.com; e.p.a.m.bakkers@tue.nl

# *List of supporting text and figures*

**Fabrication details**

**Supporting Fig1 (SI-1):** Deterministic transfer of InSb NW networks in SEM

**Supporting Fig2 (SI-2):** AFM Topography of an InSb NW network

**Supporting Fig3 (SI-3):** SEM images of all measured devices

**Supporting Fig4 (SI-4):** Evolution of conductance in magnetic field in Device-I (channels not discussed in the main manuscript)

**Supporting Fig5 (SI-5):** Voltage bias spectroscopy of Device-I (channels not discussed in the main manuscript)

**Supporting Fig6 (SI-6):** Evolution of conductance in magnetic field in Device-II

**Supporting Fig7 (SI-7):** Voltage bias spectroscopy of Device-II

**Supporting Fig8 (SI-8):** Evolution of conductance in magnetic field in Device-III

**Supporting Fig9 (SI-9):** Voltage bias spectroscopy of Device-III

# Fabrication details

- **Substrate Preparation**

  P++-doped silicon (Si) substrate with 285 nm thermal Silicon Oxide ($SiO_2$) prepared with alignment markers made of 10 nm Ti and 80 nm Au. The substrate is cleaned with acetone, isopropanol (IPA) and treated with oxygen ($O_2$) plasma.

- **Nanowires Deposition**

  InSb Nanowires(NW) networks are transferred deterministically from the growth chip onto the cleaned Si/$SiO_2$ substrate between alignment markers. The NW networks are transferred in a Zeiss scanning electron microscope (SEM) at 3 KV with a fine tungsten needle attached to a kleindiek nanomanipulator attached to the inside of the SEM chamber. The video shows a detailed description for the deterministic transfer of the networks. The growth is performed on InP <100> substrate patterned with Au droplets via electron beam lithography in an MOVPE reactor. Then Au- catalyzed InP <100> NWs grow vertically on the substrate and then kinked to the <111> B direction via controlling the Au-droplet dynamics. On top of the InP stems, Au-catalyzed InSb NWs grow epitaxially towards each other and they meet under an angle of 109.5° which is exactly the crystallographic angle between two <111> B directions.

  **SI-a** shows the as grown sample. To start the deterministic transfer of the InSb NW networks, the tungsten tip is approached slowly towards the growth sample. First, the islands anchoring the NW network are broken using the tip to release the network from the substrate, see **SI-b, c**. Afterwards, the network is picked up by the tip via van der Waals forces and the tip is moved upwards slowly and carefully to a safe distance to avoid crashing the tip or dropping the network. After making sure that the network is picked up properly and it is stable on the tip, the tip is moved to the target device chip and finally the network is placed deterministically between lithographically predefined markers as shown in **SI-d**.

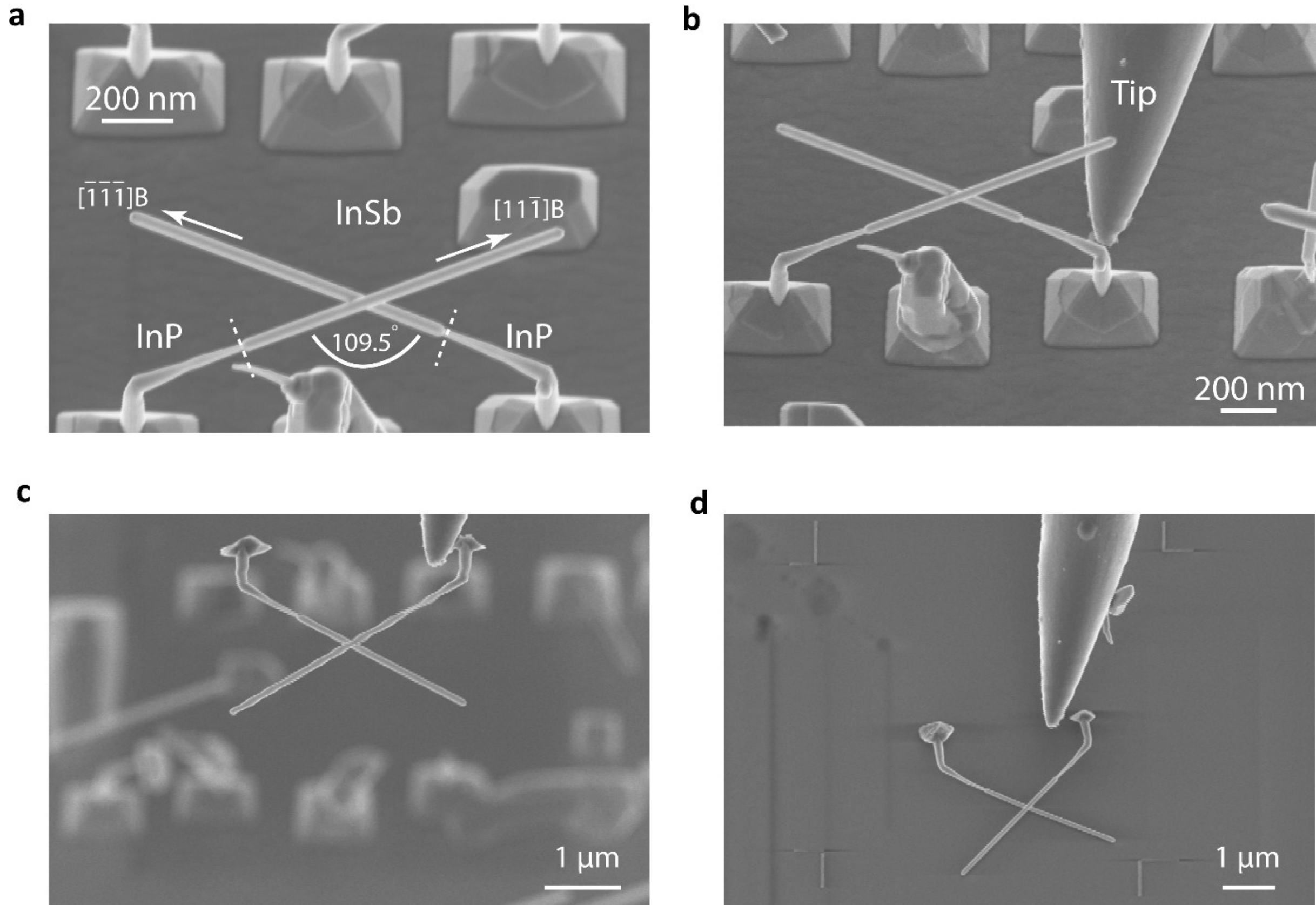

**SI-2.** Deterministic transfer of InSb NW networks in SEM (a) As grown sample of merged InSb NWs grown on InP stems. (b) A tungsten tip, controlled by a nanomanipulator, is used to transfer the NW network by breaking the islands first and picking up the network by van der Waals force as shown in (c). (d) An InSb NW network deterministically deposited between lithographically defined markers on the desired device chip (Si/SiO2 substrate). All images are 30-degree tilted SEM images.

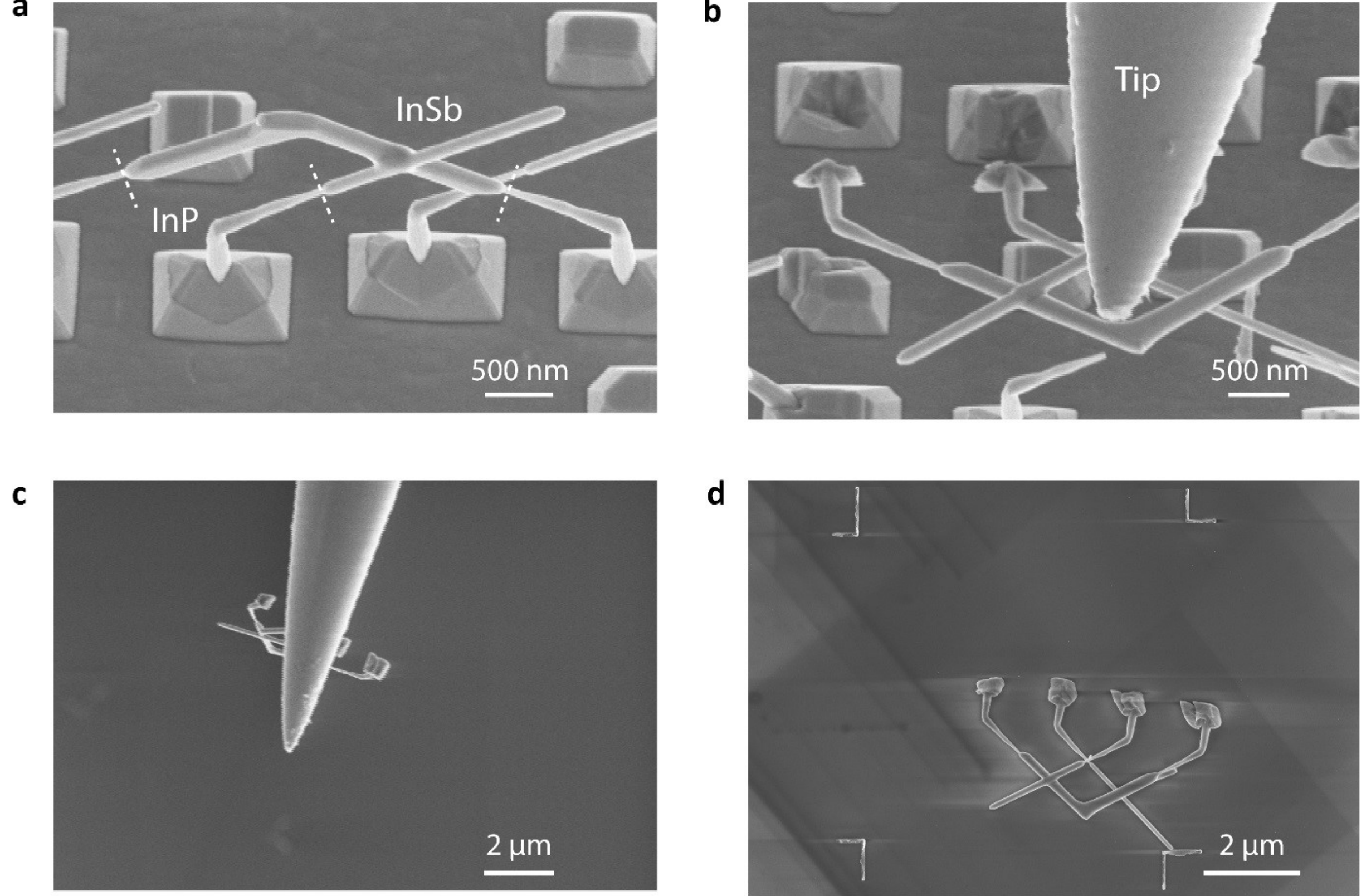

**SI-2.** Deterministic transfer of a different InSb NW network geometry than what is shown SI-1 in SEM. This network corresponds to the NW network device in the main manuscript (Device I). All images are 30-degree tilted SEM images.

- **Contact Deposition**
  - SEM image of the NW network among the predefined markers o be utilized for the contact design.
  - Resist Spinning: PMMA 495K – A6 spinned at 2000 rpm, baked at 165°C for 30 min and then PMMA 950K – A2 spinned at 2000 rpm, and baked at 165 C for 30 min.
  - Writing contact patterns using electron beam lithography
  - Developing for 60 sec in MIBK (4-Methylpentanon-2-one): IPA (Isopropanol), ratio 1:3
  - Rinsing for 60 sec in IPA and blow drying with $N_2$ gun.
  - Cleaning PMMA residues with oxygen plasma (60 sec, 1 mbar, 600 W)

- Sulphur Passivation as described in Ref.S1: Diluted ammonium polysulfide $(NH_4)_2S_x$ solution (3 ml of $(NH_4)_2S_x$ mixed with 290 mg of sulfur powder then diluted with DI-water at a ratio of (1: 200) and let to stir for 30 min at 60 °C.
- He Ion etching with a Kauffman ion source for 30 sec.
- Evaporation of 10 nm Cr and 210 nm Au.
- Lift-off in acetone overnight at room temperature.

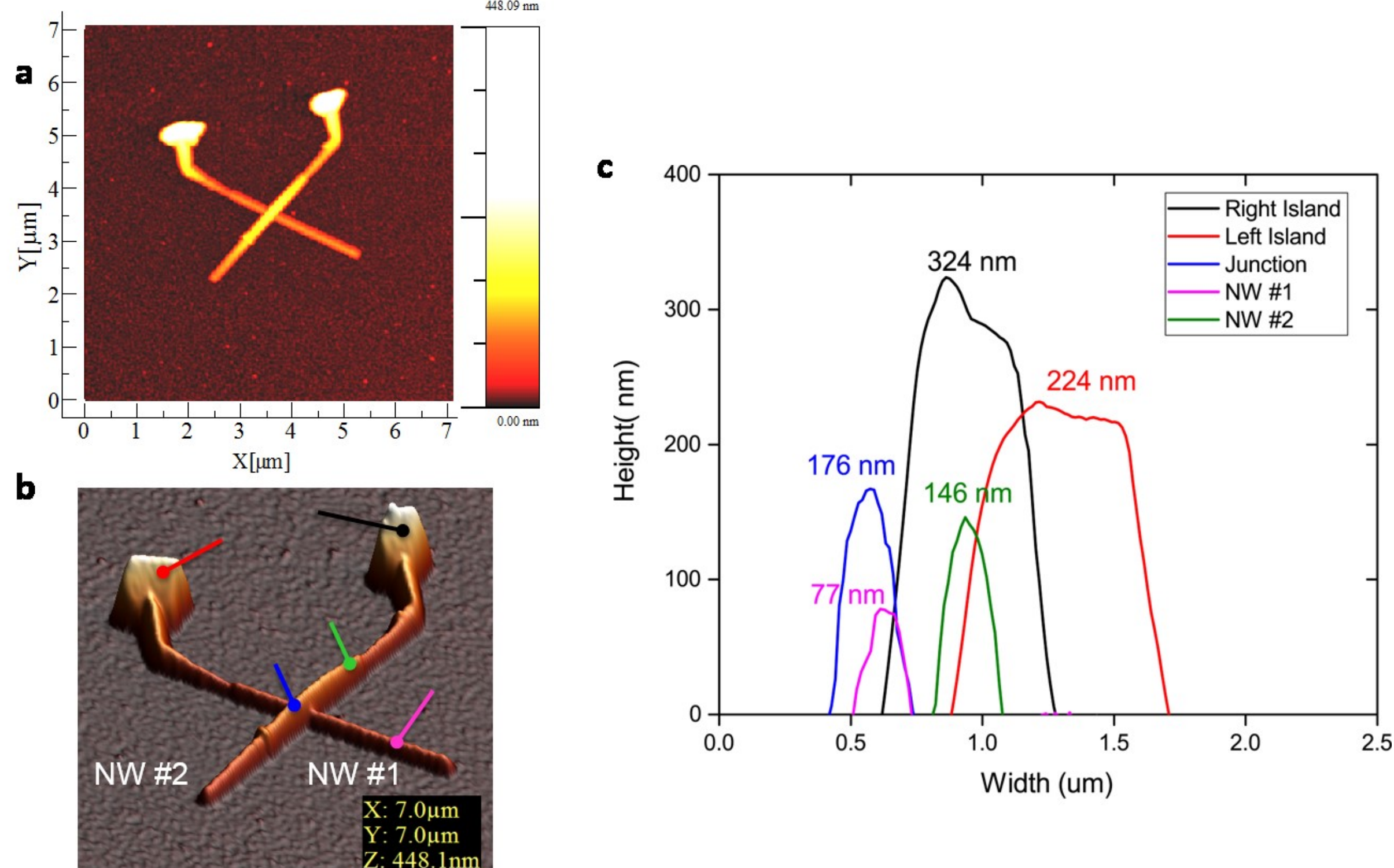

**SI-3. AFM Topography of an InSb NW network** (a) AFM image showing the height profile of a typical InSb NW network (b) 3D reconstruction of the NW network in (a). (c) height profiles of the different spots in the NW network indicated in colored arrows. The height of the islands marked in black and red are 324, 224 nm, respectively. The height of the bottom NW (NW #1) marked in (hot pink) lying on the substrate is 77 nm and the top NW (NW #2) that is lifted from the substrate from the island side (green) is 146 nm. The crossing point between the two NWs (blue) is 176 nm.

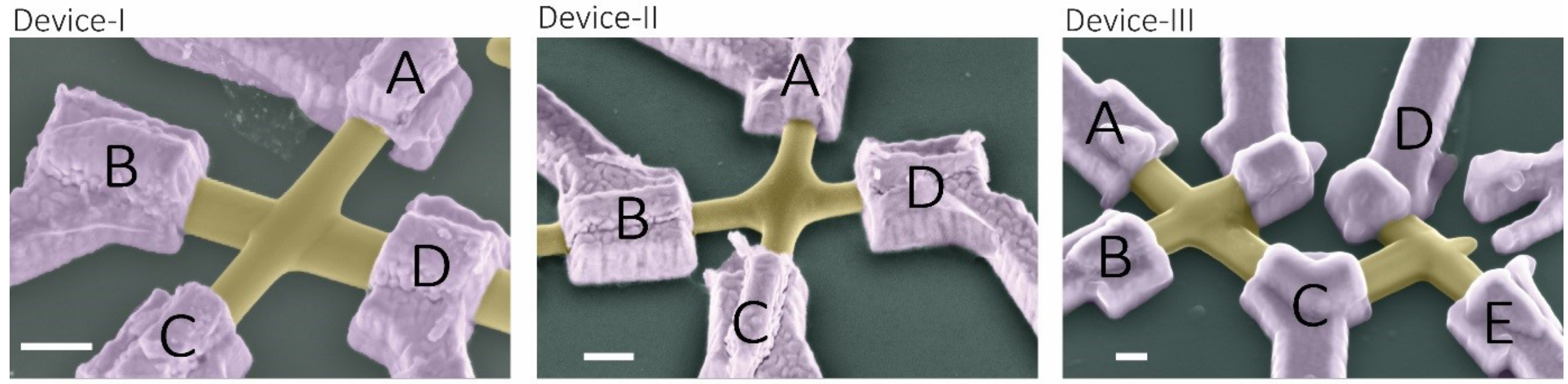

**SI-4. False color 30-degree tilted SEM images of all measured devices.** InSb NW networks are shown in yellow and the metal (Cr/Au) contacts are shown in purple. A, B, C, D, E, F are labels of the metal leads contacting the NW network. Device-I is the device discussed in the main text. Scale bars for all SEM images indicate 200 nm.

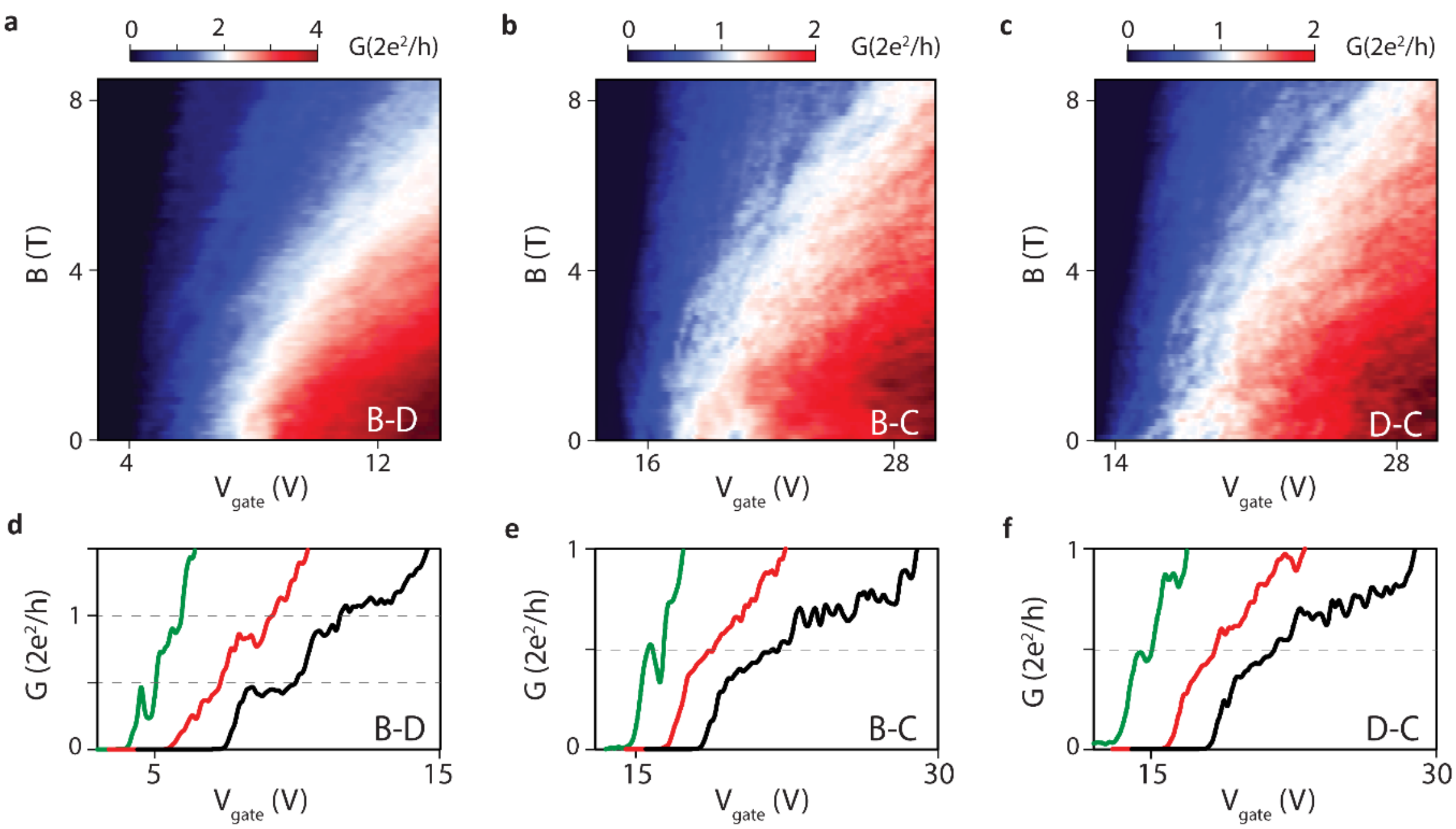

**SI 5. Evolution of conductance in magnetic field in Device-I.** This figure explains the rest of the device channels that are not discussed in the main manuscript. (a-c) Color plots of conductance $G = dI/dV_{bias}$ as a function of $V_{gate}$ and magnetic field along B at $V_{bias} = 0$ mV for different contact pair combinations: (a) B-D, (b) B-C, (c) D-C. (d-f) I-V traces indicating line cuts of (a-c) at different B values 0, 4, and 8.5 T indicated by black, red and blue, respectively; drawn with horizontal offset between the individual traces for clarity.

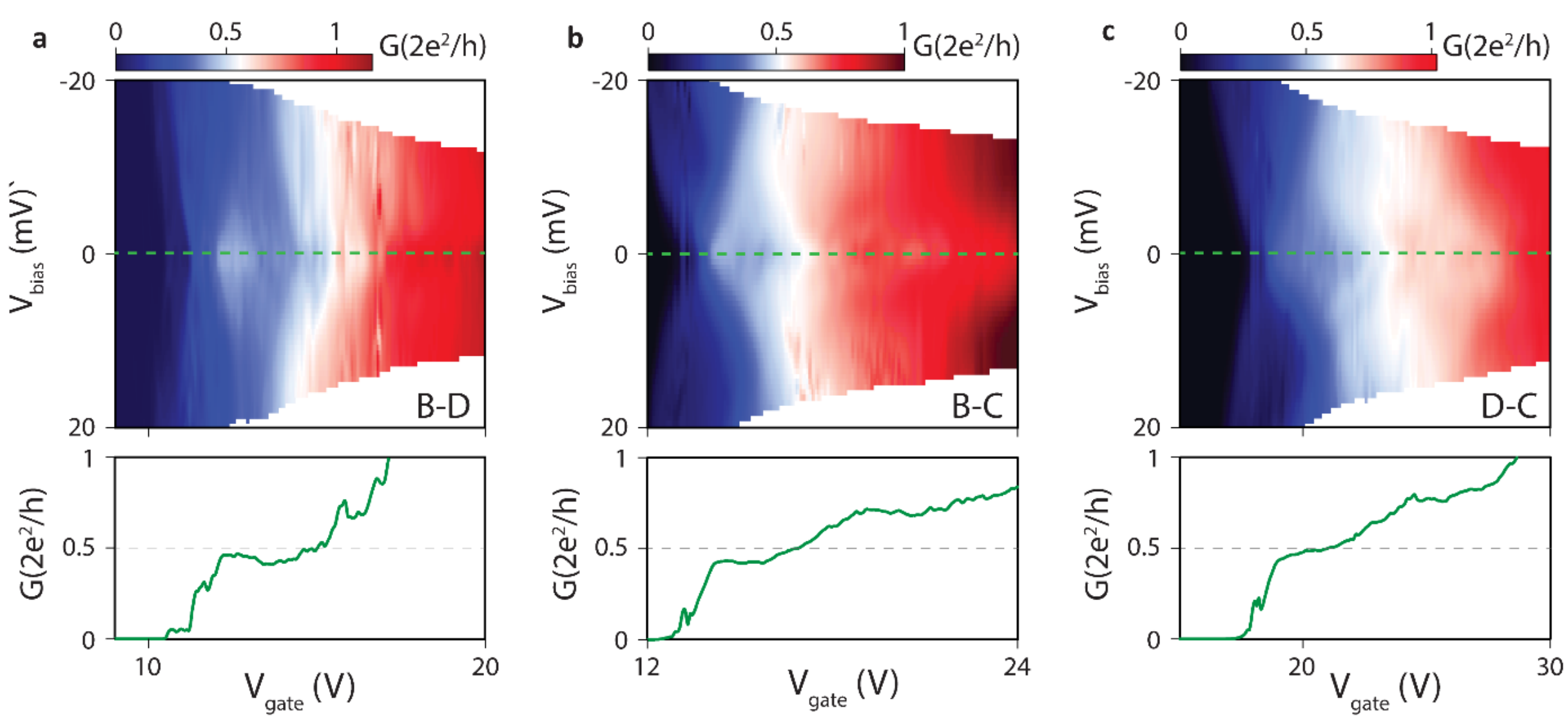

**SI 6. Voltage bias spectroscopy of Device-I.** This figure discusses the channels that are not discussed in the main manuscript. (a-c) color-plots of the differential conductance $G = dI/dV_{bias}$ as a function of $V_{bias}$ and $V_{gate}$ at $B = 8.5$ T. A line cut along $V_{bias} = 0$mV (green) is shown in the bottom panel.

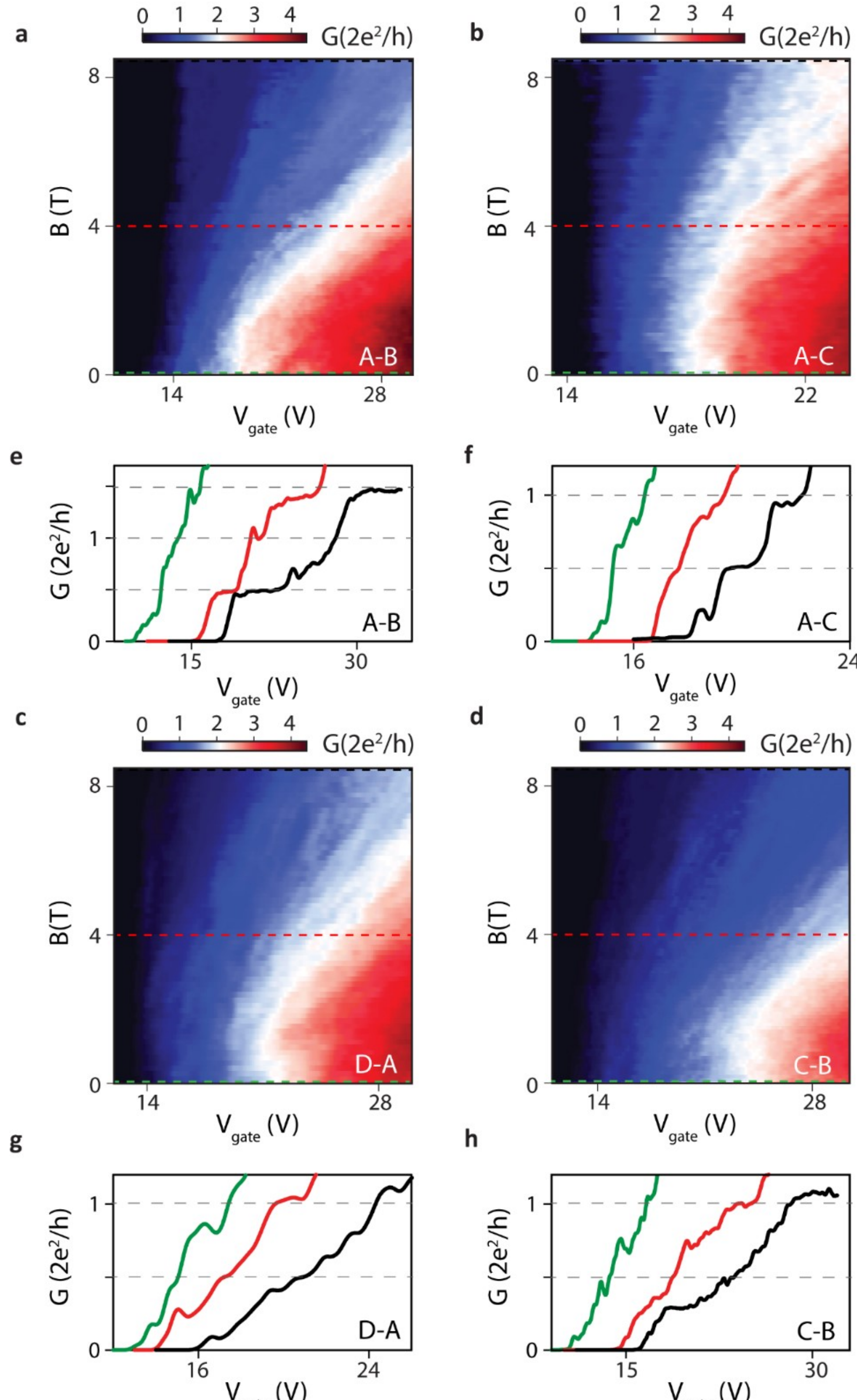

**SI-7. Evolution of conductance in magnetic field in Device-II.** (a-d) Color plots of conductance $G = dI/dV_{bias}$ as a function of $V_{gate}$ and magnetic field along B at $V_{bias}$ = 2 mV for different contact pair combinations: (a) A-B, (b) A-C, (c) D-A. (d-f) C-B. (e-h) I-V traces indicating line cuts of (a-c) at different B values 0,4, and 8.5T indicated by black, red and blue, respectively; drawn with horizontal offset between the individual traces for clarity.

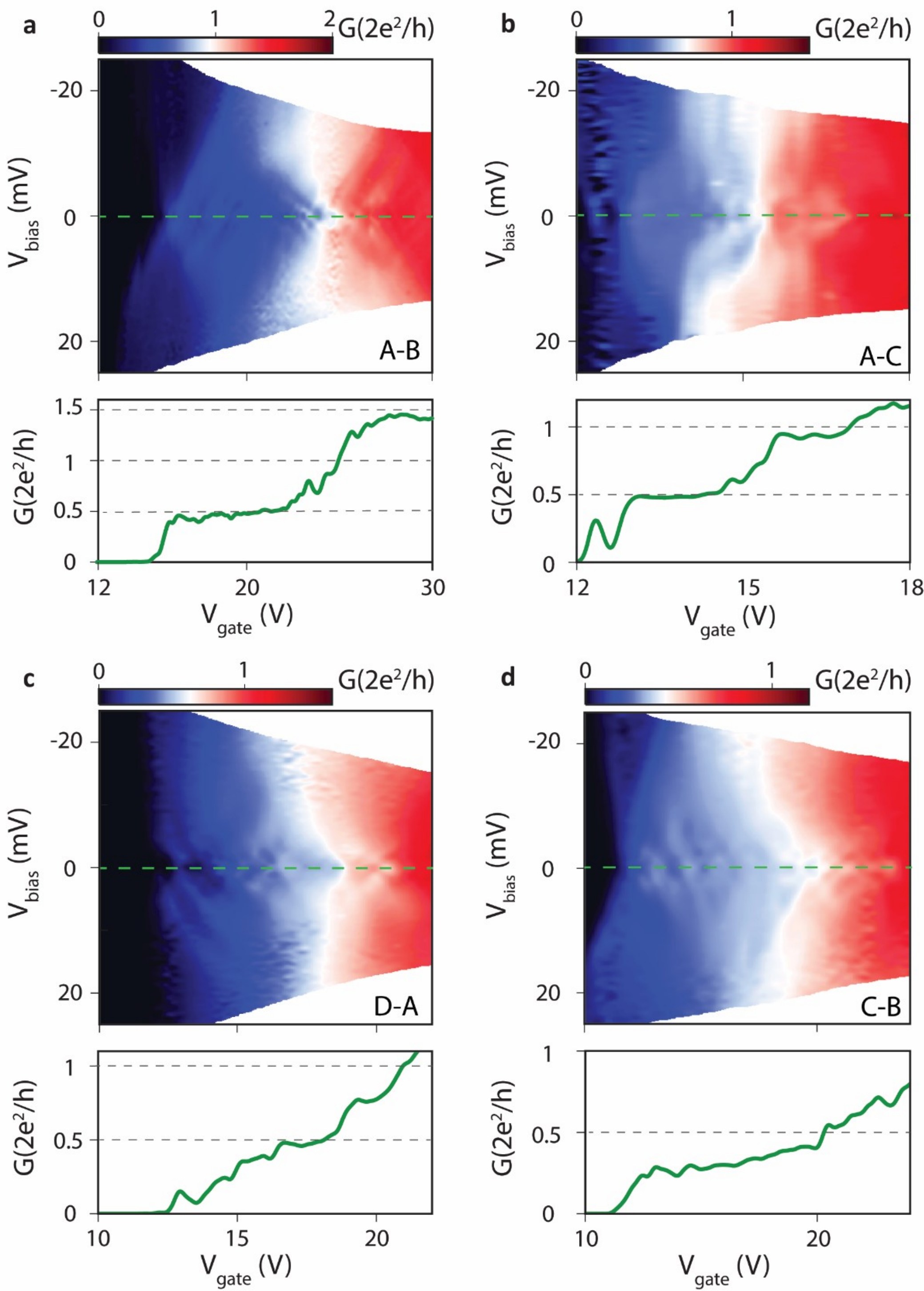

**SI-8. Voltage bias spectroscopy of Device-II.** (a-d) Color-plots of the differential conductance $G = dI/dV_{bias}$ as a function of $V_{bias}$ and $V_{gate}$ at $B = 8.5$ T. A line cut along $V_{bias} = 0$ mV (green) is shown in the bottom panel.

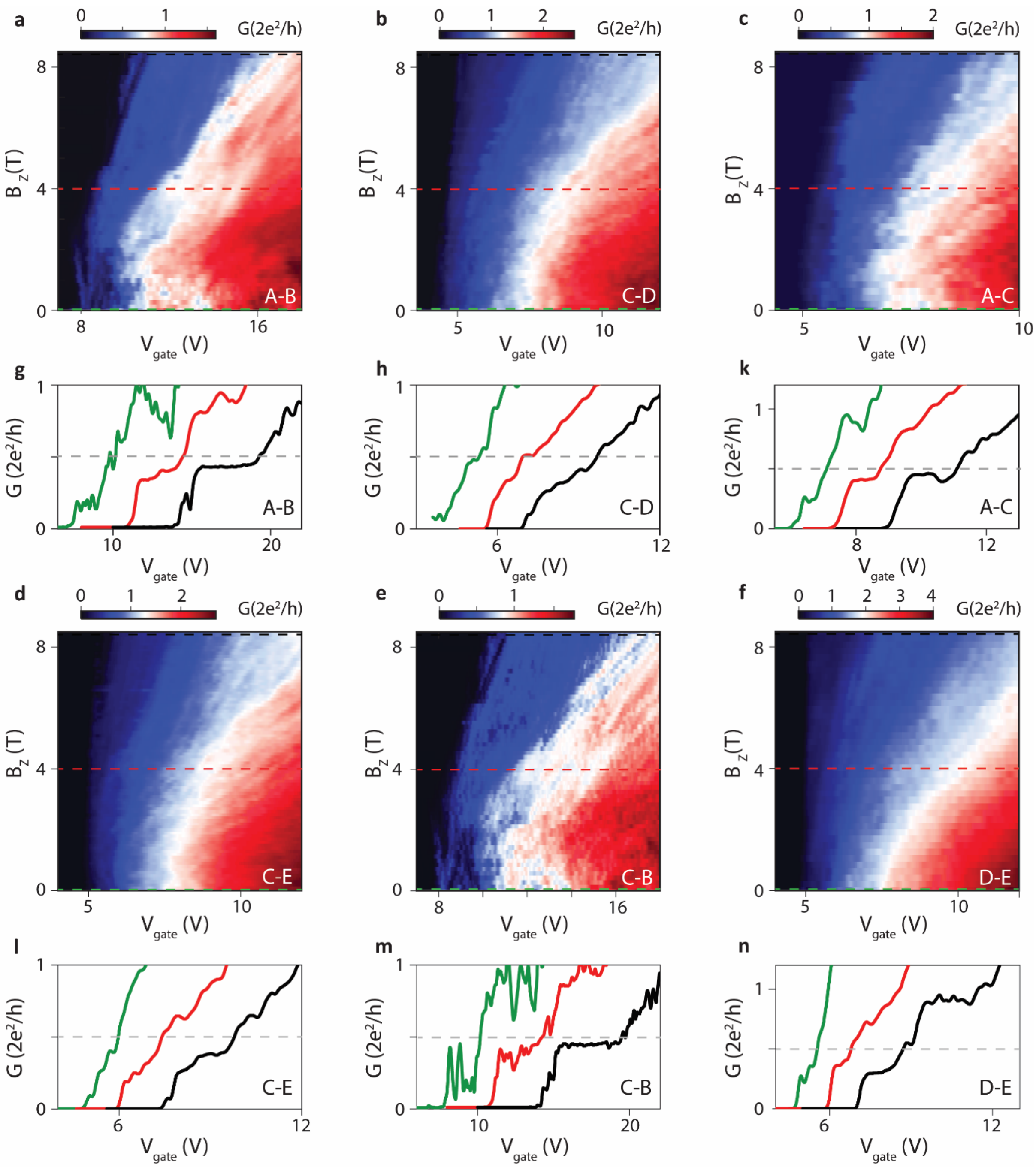

**SI-9. Evolution of conductance in magnetic field in Device-III.** (a-f) Color plots of conductance $G = dI/dV_{bias}$ as a function of $V_{gate}$ and magnetic field along B at $V_{bias} = 2$ mV for different contact pair combinations: (a) A-B, (b) C-D, (c) A-C. (d) C-E. (e) C-B and (f) D-E) (g-n) I-V traces indicating line cuts of (a-f) at different Bz values 0,4, and 8.5T indicated by black, red and blue, respectively; drawn with horizontal offset between the individual traces for clarity.

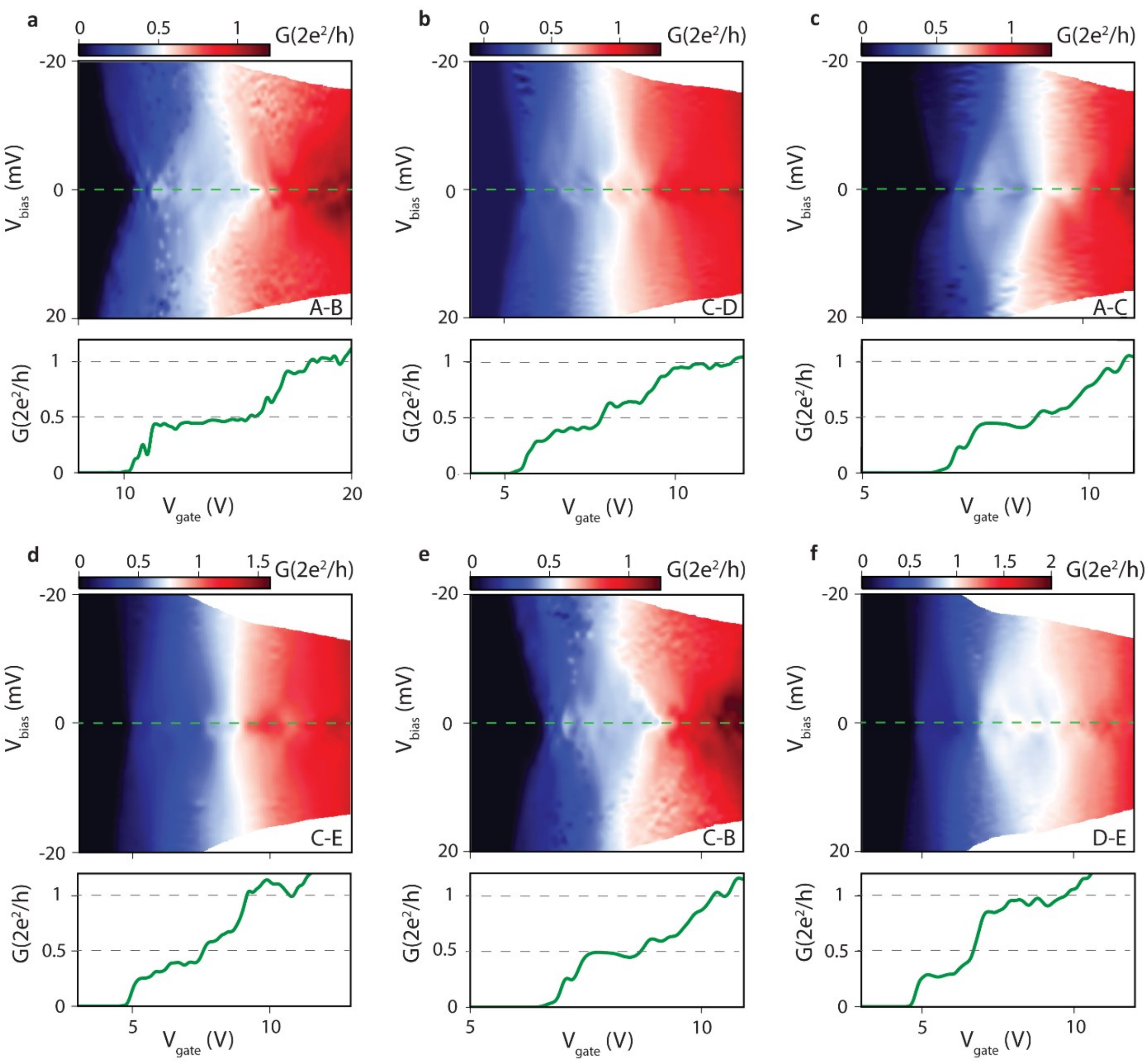

**SI-10. Voltage bias spectroscopy of Device-III.** (a-f) Color-plots of the differential conductance $G = dI/dV_{bias}$ a function of $V_{bias}$ and $V_{gate}$ at $B = 8.5$ T. A line cut along $V_{bias} = 0$ mV (green) for each color plot is shown in the bottom panel.